\documentclass[%
reprint,
superscriptaddress,
nofootinbib,
 amsmath,amssymb,
aps,
prd,
floatfix,
]{revtex4-2}

\usepackage{amssymb,latexsym,amsfonts,amsmath,amsthm,bm,bbold,tensor, mathtools}
\DeclareMathOperator{\tr}{tr}
\usepackage{hyperref}

\newcommand{\floor}[1]{\lfloor #1 \rfloor}

\begin{document}


\title{Extended framework for the hybrid Monte Carlo in lattice gauge theory}

\author{Norman H.~Christ}%
\affiliation{%
Physics Department, Columbia University, New York, NY 10027, USA
}%
\author{Lu-Chang Jin}%
\affiliation{%
Physics Department, University of Connecticut, Storrs, Connecticut 06269-3046, USA
}%
\author{Christoph Lehner}%
\affiliation{%
 Fakultät für Physik, Universität Regensburg, Universitätsstraße 31, 93040 Regensburg, Germany
}%
\author{Erik Lundstrum}%
\affiliation{%
Physics Department, Columbia University, New York, NY 10027, USA
}%
\author{Nobuyuki Matsumoto}%
 \email{nmatsum@bu.edu}
\affiliation{%
 Hariri Institute for Computing and Computational Science and Engineering, Boston University,
 Boston, MA 02215, USA
}%

\date{\today}

\begin{abstract}
We develop an extended framework
for the hybrid Monte Carlo (HMC) algorithm
in lattice gauge theory
by embedding the $SU(N)$ group
into the space of general complex matrices,
$M_N(\mathbb{C})$.
Auxiliary directions
will be completely factorized in the path integral,
and the embedding does not alter
the expectation values of the original theory.
We perform the
molecular dynamics updates
by using the matrix elements of 
$W \in M_N(\mathbb{C})$
as the dynamical variables
without group theoretic constraints.
The framework enables 
us to introduce non-separable Hamiltonians 
for the HMC in lattice gauge theory exactly, 
whose immediate application
includes the Riemannian manifold HMC.
\end{abstract}

\maketitle

\section{Introduction}
\label{sec:intro}

The hybrid Monte Carlo (HMC)
algorithm \cite{Duane:1987de} has been widely accepted as
an efficient algorithm for lattice QCD.
From the early times of its development,
strategies for curing critical slowing down
have been discussed.
The relevance of this issue is becoming increasingly high
as the field is now committed to precision calculations
using large scale machines.

One of the major promising ideas for critical slowing down
is Fourier acceleration \cite{Parisi:1984cy,Batrouni:1985jn,Davies:1989vh},
which introduces a nontrivial kinetic term in the HMC Hamiltonian
to align the effective masses for all the Fourier modes.
In generalizing the idea to gauge theory,
the kernel may be modified to
the covariant Laplacian \cite{Duane:1986fy,Duane:1988vr},
making it inevitably dependent on the gauge field,
which results in the non-separable form of the HMC Hamiltonian:
\begin{align}
  H(U,\pi) = \frac{1}{2} \pi_a G_{ab}^{-1}(U) \pi_b + S(U)
  +
  \frac{1}{2} \log \det G(U).
  \label{eq:nonsepH}
\end{align}
The degrees of freedom of the kernel matrix $G_{ab}(U)$
can be utilized to accelerate the low modes
\cite{10.1111/j.1467-9868.2010.00765.x, Nguyen:2021zgx, Jung:2024nuv},
in which context
the algorithm is dubbed Riemannian manifold (RM) HMC.

Despite its physically attractive concept,
use of the non-separable Hamiltonian~\eqref{eq:nonsepH}
in gauge theory is not straightforward.
Indeed, as was mentioned by Duane and Pendleton in Ref.~\cite{Duane:1988vr},
the discretized molecular dynamics (MD) update
generically violates the symplecticity
for a nontrivial $G(U)$.
The complication is due to the fact that,
because the gauge variable is group valued, $U(n,\mu) \in SU(N)$,
we need to exponentiate the force vector
to update $U(n,\mu)$.
Though the continuous Hamiltonian equations
preserve the symplectic two-form $\omega$,
for a finite time increment,
the exponentiation induces nonlinear terms
that generically do not cancel in the change of $\omega$
(see App.~\ref{sec:problem}).
The present paper is motivated by the recognition that the substantial
earlier work of our collaboration developing the RMHMC algorithm for QCD~\cite{Cossu:2017eys, Nguyen:2021zgx, Jung:2024nuv, Fields:2025ydm}
relied on an inexact updating algorithm which contained finite step size
errors --- a problem rectified by the extended algorithm presented here.

This paper aims to extend the framework of the HMC
in lattice gauge theory
such that 
non-separable Hamiltonians
can be used 
exactly
without gauge fixing
(see Refs.~\cite{Davies:1987vs,Sheta:2021hsd} for gauge fixing in this context).
Since the source of the issue is the constraint $U(n,\mu) \in SU(N)$,
we embed $SU(N)$ into
the space of complex $N$ by $N$
matrices, $M_N(\mathbb{C})$.
The physical $SU(N)$ variables are 
identified
by a polar decomposition.
We run the HMC for the two-dimensional $SU(3)$ pure gauge theory to test the embedding framework,
in which we take
the real and imaginary parts of the matrix elements:
$W \equiv (w_{jk}) \in M_N(\mathbb{C})$,
$w_{jk} \equiv x_{jk}+iy_{jk}$
as dynamical variables
and use the symplectic integrators
with the linear form
as in unconstrained theories.
The exactness of the algorithm is confirmed by comparing the numerical expectation values to the analytical values.

\section{Algorithm}
\label{sec:algorithm}

We use a single group variable
to demonstrate the algorithm for simplicity.
Due to the tensor product structure of the
configuration space,
generalization to the lattice gauge system is straightforward.

\subsection{Basic idea}
\label{sec:idea}

\begin{figure}[htb]
  \centering
  \includegraphics[width=5cm]{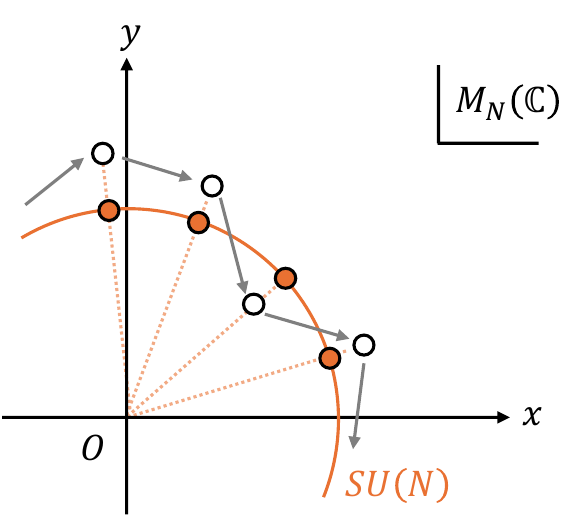}
  \caption{A schematic sketch of the extended framework.
    The MD update is performed in the $W$-space (represented by white circles),
    and the physical variable $U$ is obtained
    by using the decomposition~\eqref{eq:polar}
    as a projective map: $M_N(\mathbb{C}) \to SU(N)$
    (represented by orange, filled circles).
  }
  \label{fig:schematic}
\end{figure}

The idea is to allow the dynamical variable
in the simulation to deviate from 
$SU(N)$
into the space of complex matrices, $M_N(\mathbb{C})$ (see Fig.~\ref{fig:schematic}).
Since $M_N(\mathbb{C})$ is non-compact,
we need a systematic way to control the deviation
from $SU(N)$ to make the path integral well-defined.

For this purpose,
we parametrize the extended variable $W \in M_N(\mathbb{C})$
in the 
polar decomposition form:
\begin{align}
  W = e^{i\theta} \Phi U,
  \label{eq:polar}
\end{align}
where $\Phi$ is a positive $N\times N$ hermitian matrix and $U \in SU(N)$.
The $SU(N)$ part is the physical variable,
and the non-compact nature of $M_N(\mathbb{C})$ is fully described by $\Phi$.
As an example,
we choose the following action
for the $\Phi$ direction:
\begin{align}
  S_0(\Phi)
   \equiv
    \frac{\lambda}{2N} \,
    {\rm tr} \,
    (\Phi-\mathbb{1}_N)^2
    -
    \frac{\kappa}{N}
    \log \det \Phi
    \label{eq:s0}
\end{align}
with the tunable parameters $\lambda, \kappa >0$.
We do not add an action for the $\theta$
direction, though this is in principle possible.
The action $S_0(\Phi)$
is chosen such that
it prevents $\Phi$ 
from having an excursion to infinity
as well as from approaching the singular points,
$\det W = 0$,
at which the decomposition will be ill-defined.

Note that the value of $\Phi$ in
the decomposition~\eqref{eq:polar}
does not change under
right multiplication of a $U(N)$ matrix.
This suggests that,
for a $U(N)$-invariant measure $(dW)$,
the corresponding decomposition of the
integration measure:
\begin{align}
  (dW) = \sqrt{\det g}
  \, d\theta
  \, (d\Phi)
  \, (dU)
\end{align}
has the Jacobian
factor $\sqrt{\det g}$ that only depends on $\Phi$.
The path integral is then
completely factorized into the physical part
and the auxiliary part.
This factorized structure is reminiscent of the
gauge fixing in the path-integral.
We emphasize, however, that
we are enlarging the configuration space
for an algorithmic purpose,
and not altering the original physical system at all.

In the following,
we formulate the mathematical details of the above outline.
The resulting algorithm is simple and
is described in Sec.~\ref{sec:rmhmc_gln}.

\subsection{Decomposition along the MD trajectory}
\label{sec:polar}

We first define a decomposition of $W$ that uniquely determines
the gauge field $U$ along the MD trajectory.
An important point is that the decomposition is in accord with reversibility, which is necessary for the resulting HMC algorithm to be exact.

As is well known, for an invertible matrix $W \in GL(N, \mathbb{C})$,
the polar decomposition:
\begin{align}
  W = \Phi \Omega
  \label{eq:polar_VQ}
\end{align}
is unique, where $\Phi$ is
a positive hermitian matrix and $\Omega \in U(N)$.
We assume below
that the appearance of the singular points $\det W=0$
is properly suppressed by an appropriate choice of $\lambda$ and $\kappa$
in Eq.~\eqref{eq:s0}.

To perform
the decomposition of $\Omega \in U(N)$
into $e^{i\theta} \in U(1)$ and $U \in SU(N)$,
we need to consider an ambiguity 
in relation
to the center $\mathbb{Z}_{N}$ of $SU(N)$.
Indeed, let us define:
\begin{align}
  \theta_0 \equiv \frac{1}{N} \arg \det \Omega,
    \quad
    U_0 \equiv e^{-i\theta_0} \Omega,
    \label{eq:thetaU_0}
\end{align}
where by the argument of a complex number $z$
we intend an angle $-\pi \leq \arg z < \pi$.
Then, 
all the following 
pairs $(\theta_n, U_n)$
give
a consistent decomposition
$\Omega = e^{i \theta_n} U_n$:
\begin{align}
  \theta_n \equiv 
  \theta_0 + 2\pi n/N
  ~~({\textrm{mod }} 2\pi),
  \quad
  U_n \equiv e^{-2\pi i n/N} 
  U_0,
  \label{eq:solutions_ZN}
\end{align}
where $n = 0, \cdots, N-1$.
In other words,
multiplying $U$ 
by an 
element of the center,
$e^{2\pi i n/N}$ 
$(n=0,\cdots,N-1)$,
can be absorbed into a
shift of $\theta$
--- a shift which corresponds to the ambiguity in the definition
of the $\arg$ function in Eq.~\eqref{eq:thetaU_0}.

We perform the decomposition such that it reduces to a continuous evolution for infinitesimal updates.
Suppose we have a configuration $W$ 
and its decomposition $(U, \theta, \Phi)$,
and we 
update $W$ to $W'$.
After obtaining $\Omega'$ and $\Phi'$ from Eq.~\eqref{eq:polar_VQ},
we 
determine $U', \theta'$ as:
\begin{align}
  \delta \theta & \equiv \frac{1}{N} \arg \det \big[ e^{-i\theta} \Omega' \big],
  \label{eq:dtheta}
  \\
  \theta' &\equiv \theta + \delta \theta
  ~~({\textrm{mod }} 2\pi, \theta'\in[-\pi,\pi)),
  \\
  U' &\equiv e^{-i \theta'} \Omega'.
\end{align}
The triplet $(U',\theta',\Phi')$ is uniquely determined from $(U,\theta,\Phi)$ in a reversible manner: For a reversed process from $W'$ to $W$, where $W'$ is decomposed into the triplet $(U', \theta', \phi')$, we obtain the original $(U,\theta,\phi)$ through the same decomposition scheme
(see App.~\ref{sec:reversibility} for the detailed argument for reversibility and exactness of the algorithm).

In the above, we identified the 
$U(1)$ element of the 
$U(N)$ group that is proportional to the
identity matrix.  This corresponds to taking as the 
$U(1)$ generator 
$T_0\propto \mathbb{1}_N$. 
However, other choices are possible.  Indeed, by choosing a $U(1)$ subgroup whose generator is not the identity
matrix, we can decompose 
$W$ into 
$(U, \theta, \Phi)$
without ambiguity as is worked out in App.~\ref{sec:another_decomp}.

\subsection{Path integral in the larger space}
\label{sec:path_integral}

We next consider the integration measure for $M_N(\mathbb{C})$
and its decomposition corresponding to Eq.~\eqref{eq:polar}.
The desired measure can be
conveniently defined from
the metric tensor $\textbf{g}$ as:
\begin{align}
  &\textbf{g}
  \equiv
  {\rm tr}\, [dW dW^\dagger]
  =
  \sum_{j,k}[dx_{jk}^2 + dy_{jk}^2],
  \label{eq:metric_tensor}
  \\
  &(dW)
  \equiv
  \prod_{j,k}
  \big(
  dx_{jk} \, dy_{jk}
  \big),
  \label{eq:measure_flat}
\end{align}
where the bilinear form
is assumed to be symmetrized.

To discuss the decomposition of $(dW)$,
we prepare the one-form basis for the variables
$U$ and $\Phi$.
A convenient choice for $U$
is given by the Maurer-Cartan form:
\begin{align}
  \Theta
  \equiv dU U^{-1}
  \equiv i T_a \Theta_a
  .
  \label{eq:MaurerCartan}
\end{align}
$\Theta_a$ is dual to the right-invariant derivative $D_a$:
\begin{align}
  D_a U = i T_a U,
  \quad
  dU = \Theta_a D_a U.
\end{align}
With the traceless hermitian generators $T_a$,
we expand the hermitian matrices as:
\begin{align}
  \Phi = \phi_a T_a + \phi_0 \mathbb{1}_N,
  \quad
  {\rm tr}\,[T_a T_b] = \delta_{ab}.
\end{align}
From the relation:
\begin{align}
  dW
  &=
  i e^{i\theta} \Phi T_a U \Theta_a
  +
  i e^{i\theta} \Phi U d\theta 
  \nonumber\\
  &~~~~~~~~~~~~~~
  +
  e^{i\theta} T_a U d\phi_a
  +
  e^{i\theta} U d\phi_0,
  \label{eq:expand_dw}
\end{align}
the metric tensor can be rewritten accordingly:
\begin{align}
  \textbf{g}
  &=
  {\rm tr}\, [T_{a} T_{b} \Phi^2] \, \Theta_a \Theta_b
  +
  2 \, {\rm tr}\,[T_a \Phi^2] \, \Theta_a d\theta 
  \nonumber\\
  &~~~~~~~~~~~~~~
  +
  {\rm tr}\, \Phi^2  d\theta^2
  +
  d\phi_a^2
  +
  d\phi_0^2.
\end{align}
As advertised,
the Jacobian depends only on $\Phi$:
\begin{align}
  (dW)
  =
  \sqrt{\det g(\Phi)}
  \,
  d\theta
  \,
  (dU)
  \,
  (d\Phi),
  \label{eq:measure_w}
\end{align}
where
\begin{align}
  (dU) \equiv \prod_a \Theta_a
\end{align}
is the Haar measure, $(d\Phi) \equiv d\phi_0 \,\prod_a d\phi_a$,
and
\begin{align}
  \det g(\Phi)
  & =
    \det \left[
    \begin{array}{c c}
      (1/2){\rm tr}\,[\{T_{a}, T_{b}\} \Phi^2] & {\rm tr}\,[T_b \Phi^2] \\
      {\rm tr}\,[T_a \Phi^2] & {\rm tr}\, \Phi^2
    \end{array}
    \right].
\end{align}

Collecting the formulas,
the path integral in the larger space:
\begin{align}
  Z_{M_N(\mathbb{C})}
  \equiv
  \int (dW)\,
  e^{ - S(U)- S_0(\Phi)}
  \label{eq:large_path_int}
\end{align}
can be related to the original one:
\begin{align}
  Z_{SU(N)}
  \equiv
  \int(dU)\,
  e^{-S(U)}
\label{eq:original_path_integral}
\end{align}
as follows:
\begin{widetext}
\begin{align}
  Z_{M_N(\mathbb{C})}
  &=
    \int d\theta
    \int
    (dU)\, e^{-S(U)}
    \int_{\Phi > 0}
    (d\Phi)
    \sqrt{\det g(\Phi)}
    e^{-S_0(\Phi)}
  =
    (2\pi)
    Z_{SU(N)}
    \int_{\Phi > 0}
    (d\Phi)
    \sqrt{\det g(\Phi)}
    e^{-S_0(\Phi)},
    \label{eq:Z_rel}
\end{align}
where the integration domain
${\Phi > 0}$ is over positive 
hermitian matrices.
The factorized $\Phi$ integral
is well-defined
because of the bound:
\begin{align}
  \int_{{\Phi > 0}}
  (d\Phi)
  \sqrt{\det g(\Phi)}
  e^{-S_0(\Phi)}
  <
  \int
  (d\Phi)\,
  e^{-\frac{\lambda}{2N}\tr (\Phi-1)^2}
  \sqrt{\det g(\Phi)}
  \,
  \vert \det \Phi \vert^{\frac{\kappa}{N}},
  \label{eq:gauss_phi}
\end{align}
\end{widetext}
which is a Gaussian integral
of a function that has at most a power law increase for large $\Phi$,
and thus has a finite value.

The above shows that we can calculate the expectation value of
the observable ${\cal O}(U)$ directly in the larger path integral
as a function of the physical part $U$.
The nontrivial Jacobian factor will automatically drop out
in the expectation values,
and its precise form is irrelevant for running the simulation.

\subsection{The HMC in the extended space}
\label{sec:rmhmc_gln}

We now have
a path integral~\eqref{eq:large_path_int}
over $2N^2$ unconstrained real variables
$(w_I) \equiv (x_{jk}, y_{jk})$
with the flat measure~\eqref{eq:measure_flat}.
We are therefore
ready to run the HMC for the flat space
to simulate the gauge system.

We write the momentum as $p_I$
conjugate to $w_I$.
The kinetic term of the HMC Hamiltonian
can be taken arbitrarily:
\begin{align}
  H(w,p)
  & \equiv
  \frac{1}{2} p_I K^{-1}_{IJ}(w) p_J + S(U)
  \nonumber\\
  &~~~~~~~
  + 
  S_0(\Phi)
  +
  \frac{1}{2} \log \det K(w).
  \label{eq:hamiltonian}
\end{align}
With the implicit leapfrog integrator, for example,
the entire algorithm will be the following:
\begin{enumerate}
\item Suppose we have a configuration $W$.
\item Generate $p$ from the Gaussian distribution:
  \begin{align}
    P_{\rm init}(p; w) \propto e^{-\frac{1}{2} p_I K^{-1}_{IJ}(w) p_J}.
  \end{align}
\item
  Integrate the Hamiltonian equations:
  \begin{align}
    &p_I^{1/2}
      =
      p_I
      -
      \frac{\tau}{2}
      \partial_{w_I} H(w,p^{1/2})
      ,
      \label{eq:p_half}
    \\
    &w_I^{1/2}
      =
      w_I
      +
      \frac{\tau}{2}
      \partial_{p_I} H(w,p^{1/2})
      ,
      \label{eq:q_half}
    \\
    &w_I^{\prime}
      =
      w_I^{1/2}
      +
      \frac{\tau}{2}
      \partial_{p_I} H(w',p^{1/2})
      ,
      \label{eq:q_prime}
    \\
    &p'_I
      = p_I^{1/2}
      -
      \frac{\tau}{2}
      \partial_{w_I} H(w',p^{1/2}).
      \label{eq:p_prime}
  \end{align}
\item Accept/reject the obtained configuration with the probability:
  \begin{align}
    \min \Big(
    1, e^{-H(w',p') + H(w,p)}
    \Big).
  \end{align}
\item We add to the ensemble
the physical configuration $U$,
calculated from the accepted $W$,
to estimate the 
expectation values of the observables
under the path integral~\eqref{eq:original_path_integral}.
\end{enumerate}
The symplecticity of the integrator (see App.~\ref{sec:implicit_symp})
implies the conservation
of the phase-space volume.
Together with reversibility, the exactness of the algorithm follows.

Since the action is written in the $(U, \theta, \Phi)$-basis
while the update is in the $W$-basis,
we need the Jacobian matrix
to relate the two
in the force calculation.
From Eq.~\eqref{eq:expand_dw}:
\begin{align}
  &\left[
    \begin{array}{c c}
      dx_{jk}
      &
        dy_{jk}
    \end{array}
    \right]
    =
    \left[
    \begin{array}{c c c c}
      \Theta_a & d\theta & d\phi_a & d\phi_0
    \end{array}
    \right]
    \times 
    \nonumber\\
    &~~~~~~~~~~~~~~~~
    \times 
    \left[
    \begin{array}{c c}
      - {\rm Im}\,[e^{i \theta} (\Phi T_a U)_{jk}]
      &
        {\rm Re}\,[e^{i \theta} (\Phi T_a U)_{jk}]
      \\
      - {\rm Im}\,[e^{i \theta} (\Phi U)_{jk}]
      &
        {\rm Re}\,[e^{i \theta} (\Phi U)_{jk}]
      \\
      {\rm Re}\,[e^{i \theta} (T_a U)_{jk}]
      &
        {\rm Im}\,[e^{i \theta} (T_a U)_{jk}]
      \\
      {\rm Re}\,[e^{i \theta} U_{jk}]
      &
        {\rm Im}\,[e^{i \theta} U_{jk}]
    \end{array}
    \right]
    \nonumber\\
  &
    ~~~~~~~~~~~~~~~~~~
    \equiv
    \left[
    \begin{array}{c c c c}
      \Theta_a & d\theta & d\phi_a & d\phi_0
    \end{array}
    \right]
    \times
    J(U,\theta, \Phi).
    \label{eq:transf}
\end{align}
The force in the $W$-basis
can be calculated from the $(U, \theta, \Phi)$-basis as:
\begin{align}
  \left[
  \begin{array}{c}
    \partial_{x_{jk}} H(p, w)
    \\
    \partial_{y_{jk}} H(p, w)
  \end{array}
  \right]
  =
  J(U, \theta, \Phi)^{-1}
  \left[
  \begin{array}{c}
    D_a H(p, w)
    \\
    \partial_\theta H(p, w)
    \\
    \partial_{\phi_a} H(p, w)
    \\
    \partial_{\phi_0} H(p, w)
  \end{array}
  \right].
  \label{eq:force_prop}
\end{align}

We comment that, because of the center $\mathbb{Z}_N$, we formally regard the triplet $(U,\theta,\Phi)$ as a state of the Markov chain, and the global structure of the state space is isomorphic to $GL(N,\mathbb{C})\times \mathbb{Z}_N$.
We emphasize, however, that the center $\mathbb{Z}_N$ does not appear explicitly during the HMC, and we do not need to set up an additional Monte Carlo update for the center.
The only occasions that require care are the initialization of the variables and saving a configuration to a file.
To implement a hot start,
for example,
one can randomly generate $W$,
perform the decomposition into $(U_0, \theta_0, \Phi)$
by using Eq.~\eqref{eq:thetaU_0},
and shift $(U_0, \theta_0)$
with a randomly 
chosen $n \in \mathbb{Z}_N$: 
$(U_0, \theta_0) \to (U_n, \theta_n)$
as Eq.~\eqref{eq:solutions_ZN}.
We associate the triplet $(U_n, \theta_n, \Phi)$ with the starting configuration $W$.
To store a state of the Markov chain, we save the triplet $(U,\theta,\Phi)$, from which the $W$-variable can be easily reconstructed.

The above argument for the single variable can be readily generalized for the lattice gauge system in arbitrary dimensions. 
During the simulation, we keep $W(n,\mu)$ and the triplet $(U(n,\mu),\theta(n,\mu), \Phi(n,\mu))$ for all the links $(n,\mu)$.
The Jacobian matrix that maps the gradients in the two bases is block-diagonal in the spatial and Lorentz structures.
Therefore, we can calculate the force vector for $W(n,\mu)$ by the same formula as Eq.~\eqref{eq:force_prop} by adding the link labels $(n,\mu)$.

\section{Numerical test}
\label{sec:numerical_test}

To numerically verify the exactness of our algorithm,
we perform Monte Carlo calculations
for the two-dimensional pure $SU(N = 3)$ gauge theory
using the Wilson gauge action
with $\beta=1.00,1.05,\cdots,4.00$.
We adopt the
trivial kernel $K(w)=\mathbb{1}_{2N^2}$ to test that the HMC gives exact results under the proposed embedding framework, examining it independently from additional complications from implicit iteration.

\begin{figure*}[htb]
  \centering
  \includegraphics[width=6cm]{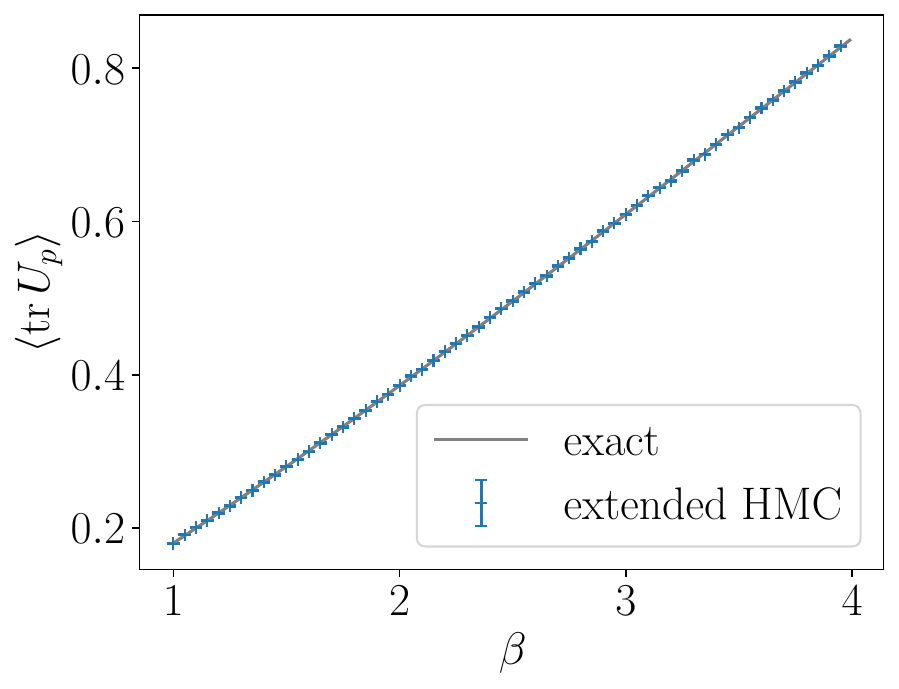}
  \hspace{1cm}
  \includegraphics[width=6cm]{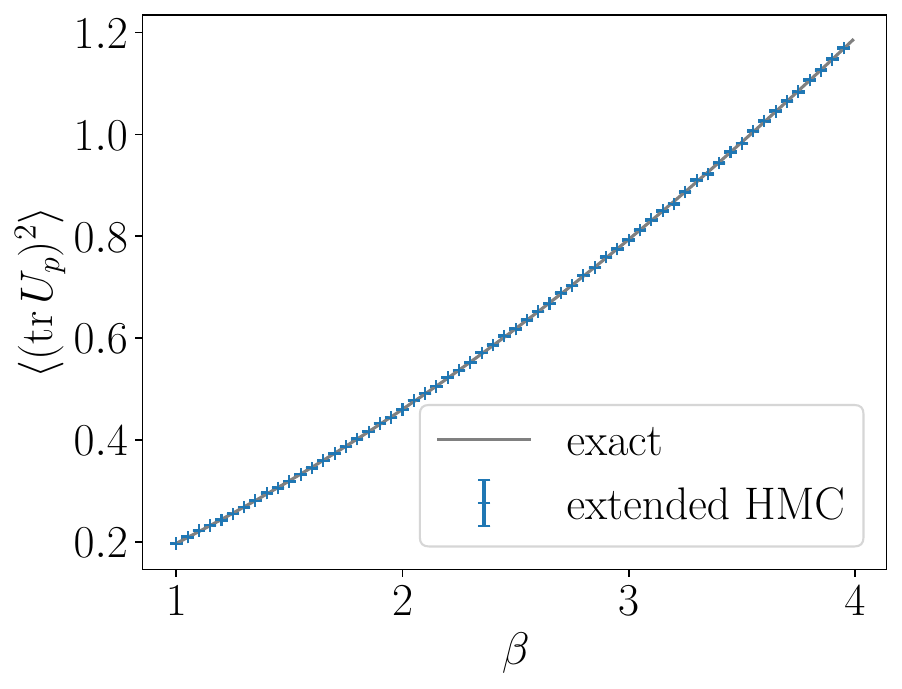}
  \caption{Expectation values of the plaquette (left)
    and the plaquette squared (right) calculated with the proposed,
    extended HMC algorithm.
    }
  \label{fig:result}
\end{figure*}

\begin{figure*}[htb]
  \centering
  \includegraphics[width=6cm]{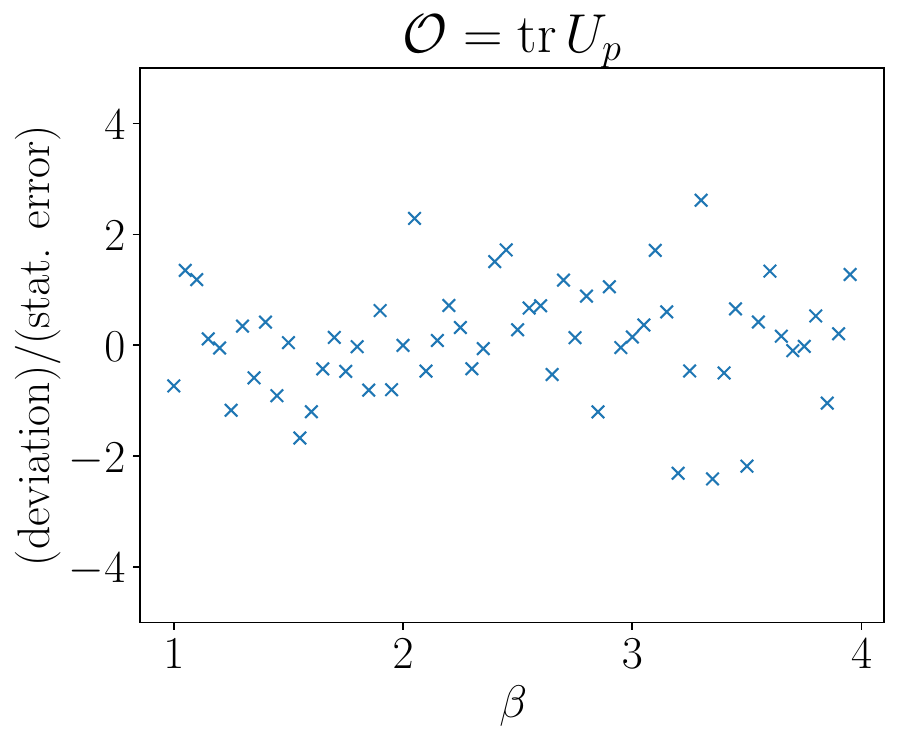}
  \hspace{1cm}
  \includegraphics[width=6cm]{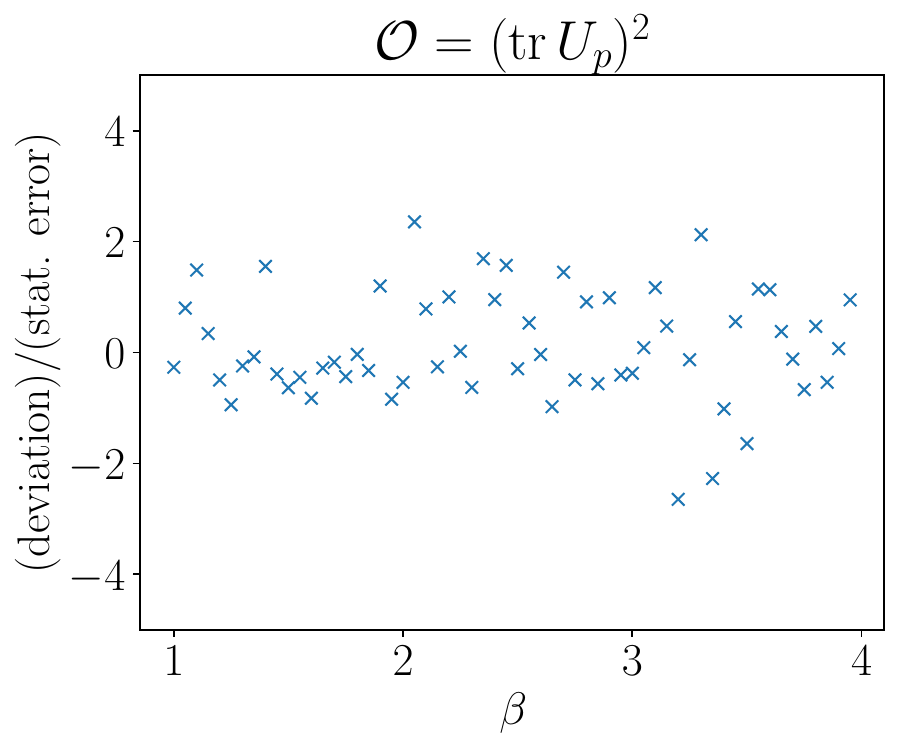}
  \caption{
    The deviation
    of the Monte Carlo estimate $\bar{\cal O}$
    from the exact value $\langle {\cal O} \rangle$
    in units of the statistical error $\delta \bar{\cal O}$:
    $(\bar{\cal O} - \langle {\cal O} \rangle) / \delta \bar{\cal O}$.
    In the left panel, ${\cal O} = {\rm tr}\, U_p$,
    and in the right, ${\cal O} = ({\rm tr}\, U_p)^2$.
    The plots show reasonable accuracy for our calculation.
  }
  \label{fig:err}
\end{figure*}

The lattice is periodic and has the dimension $32\times 32$.
The auxiliary parameters are set to
$\lambda = 1$ and $\kappa=5$.
The MD is integrated
with the explicit leapfrog.
A trajectory of length 1.0 in units of MD time
is separated into 10 steps,
which gives an acceptance around 0.8.
After thermalization from a hot start,
we calculate the observables every 10 trajectories.
The expectation values are estimated from 
1,000 configurations,
and the
statistical errors are estimated with the jackknife method.
Figure~\ref{fig:result} shows the
expectation values of the
plaquette $\langle {\rm tr}\, U_{p} \rangle$
and the plaquette squared
$\langle ({\rm tr}\, U_{p})^2 \rangle$.
The exact values are calculated from the character expansion.
The precision is sub-percent level
and all the estimates
are consistent with the exact value
within $3 \sigma$ confidence level (see Fig.~\ref{fig:err}).
Good agreement verifies the exactness of the proposed algorithm.

\section{Discussion}
\label{sec:discussion}

In this paper,
we develop a framework to extend the HMC in lattice gauge theory
by embedding $SU(N)$
into the space of general complex matrices, $M_{N}(\mathbb{C})$.
Our extension 
not only 
allows us to
use the HMC with non-separable Hamiltonians
but 
also adds new dimensions
to optimize the HMC.
For example, let us expand the momentum in the
$(U, \theta, \Phi)$-basis
in terms of 
the momentum $p \equiv (p_I)$ in the $W$-basis:
\begin{align}
 (\pi_a, \pi_\theta, \rho_a, \rho_0)^T
  \equiv 
  J(w) 
  \cdot p.
  \label{eq:p2pi}
\end{align}
An extension from the
RMHMC Hamiltonian~\eqref{eq:nonsepH} is:
\begin{align}
  H(w,p)
  &\equiv
  \frac{1}{2}
  \pi_a G^{-1}_{ab} (U) \pi_b
  +
  \frac{1}{2 m_\theta}
  \pi_\theta^2
  +
  \frac{1}{2 m_{\rho}}
  \rho_a^2
  +
  \frac{1}{2 m_0}
  \rho_0^2 
  \nonumber\\
  &~~~~~~~~
  +
  S(U)
  +
  S_0(\Phi)
  +
  \frac{1}{2} \log \det G(U).
  \label{eq:block_diag_kernel}
\end{align}
As shown in App.~\ref{sec:reduced},
the continuous 
Hamiltonian equations
for the Hamiltonian~\eqref{eq:block_diag_kernel}
reduce to those of the
original RMHMC
when the evolution is projected
onto $SU(N)$.  (Of course, if the Hamiltonian in Eq.~\eqref{eq:block_diag_kernel} is to be used in the symplectic finite-time-step update given in Eqs.~\eqref{eq:p_half}--\eqref{eq:p_prime}, the momenta $(\pi_a, \pi_\theta, \rho_a, \rho_0)$ must first be expressed in terms of the $p_I$ using Eq.~\eqref{eq:p2pi}.) 
However,
it is 
possible 
in principle to 
intentionally mix the physical modes $\pi_a$
with the auxiliary modes $\pi_\theta$, $\rho_a$ and $\rho_0$
by adding off-diagonal terms.
It is interesting
to consider if the additional
degrees of freedom can be utilized to
increase the tunneling
rate of the topological charge.
Machine learning
may be especially useful for 
this purpose
(see Refs.~\cite{pmlr-v97-cohen19d,Kanwar:2020xzo,Favoni:2020reg,Lehner:2023bba,Lehner:2023prf,Nagai:2023fxt}
for gauge invariant
neural networks).

Avoiding the singular points,
at which $\det W = 0$,
is crucial in practice.
In the vicinity of the singular points,
$\Phi$ can take arbitrarily large values,
resulting in a slow convergence
in the
decomposition~\eqref{eq:polar}
and a large gradient $\partial S_0(\Phi)$.
In this regard,
a detailed study of the optimal choice of $S_0(\Phi)$
is important.
In our simple
choice of the action,
Eq.~\eqref{eq:s0},
large $\lambda$ 
restricts the fluctuation in
the non-compact directions to be small,
while it can also 
cause an unbalance between
the forces from $S(U)$ and $S_0(\Phi)$.
In the RMHMC applications,
therefore,
it is important to
adjust the relative sizes of the canonical masses, $m_\theta$, $m_\rho$ and $m_0$ in Eq.~\eqref{eq:block_diag_kernel},
at the same time
when we engineer the 
auxiliary action $S_0$.
It may be also beneficial
if we can utilize
the multilevel integration scheme \cite{SEXTON1992665}
by separating the physical and auxiliary actions.
We note that
the effective action of $\Phi$
has a nontrivial contribution
from the measure 
[see Eq.~\eqref{eq:Z_rel}],
which can be understood as the 
potential term for a centrifugal force
from the singularity
(see Fig.~\ref{fig:schematic}).
It may be important
to design $S_0(\Phi)$ taking this effect into
account.
In our exploratory runs in 
four-dimensions,
we find that
using a quartic 
instead of a quadratic form
can increase the efficiency.

Finally, since our goal is to speed up lattice QCD production runs, it is most important to test the algorithm in realistic systems.
Though the iterations in the implicit update steps are potentially computationally demanding, it is important to recognize that the increase of the computational cost from these steps is additive to the fermion force calculation in the usual multilevel evolution algorithm~\cite{SEXTON1992665}.
Indeed, we can circumvent reevaluating the fermion force in the implicit iteration by pushing the force terms for the nontrivial kernel $K(w)$ into the lower level where we calculate the gauge force
\cite{Nguyen:2021zgx,Jung:2024nuv}.
Consequently, the net increase in computational cost can be mild for the dynamical simulations where the cost for the fermion force is substantial.
Especially under such circumstances,
combining the algorithm
with the field-transformation \cite{Luscher:2009eq}
may yield further benefit.

Work along these lines is in progress
and will be reported elsewhere.

\section*{Acknowledgments}
The authors thank Peter Boyle,
Richard C.~Brower, 
Sarah Fields,
Taku Izubuchi,
Chulwoo Jung 
and
Joseph V.~Pusztay
for valuable discussions.
The authors further express gratitude to referee of Physical Review D for providing thoughtful comments that helped revise the manuscript significantly.
This work is supported by 
the Scientific Discovery through Advanced Computing (SciDAC) program, 
``Multiscale acceleration: Powering future discoveries in High Energy Physics'' under FOA LAB-2580
funded by U.S. Department of Energy (DOE), Office of Science.
L.C.J. acknowledges support by DOE Office of Science Early Career
Award No.~DE-SC0021147 and DOE Award No.~DE-SC0010339. N.H.C. and E.L. 
are
supported in part by U.S. 
DOE
grant No.~DE-SC0011941,
and N.M. in part by U.S. 
DOE
grant No.~DE-SC0015845.

\appendix

\section{Comment on implicit, \\ 
symplectic integrators}
\label{sec:implicit_symp}

The implicit leapfrog integrator,
Eqs.~\eqref{eq:p_half}--\eqref{eq:p_prime},
is a symplectic discretization of the Hamiltonian equations:
\begin{align}
  \dot{w}_I = \partial_{p_I} H,
  \quad
  \dot{p}_I = - \partial_{w_I} H,
\end{align}
which are derived using the symplectic two-form:
\begin{align}
  \omega(w,p) \equiv
  dp_I \wedge dw_I.
\end{align}
Indeed, as one can easily show,
for the first pair of the update,
$\omega$ is preserved:
\begin{align}
  &\omega(w^{1/2},p^{1/2}) \nonumber\\
  &=
  \Big(
  \delta_{IJ} +
  \frac{\tau}{2}
  \partial_{p_I}
  \partial_{w_J}
  H(w,p^{1/2})
  \Big)
  dp_I^{1/2}\wedge dw_J \nonumber\\
  &=
  \omega(w,p).
  \label{eq:conservation_unconstrained}
\end{align}
The second pair
is determined from the time-reversal symmetry.

This structure may become more transparent by
introducing the auxiliary transfer matrices:
\begin{align}
  T^{p,\tau/2}
  \left(
  \begin{array}{c}
    w \\
    p
  \end{array}
  \right)
  =
  \left(
  \begin{array}{c}
    w \\
    p - \frac{\tau}{2} \partial_{w}H(w,p)
  \end{array}
  \right),
  \\
  T^{q,\tau/2}
  \left(
  \begin{array}{c}
    w \\
    p
  \end{array}
  \right)
  =
  \left(
  \begin{array}{c}
    w + \frac{\tau}{2} \partial_{p}H(w,p) \\
    p
  \end{array}
  \right),
\end{align}
which represent explicit steps.
The process of incrementing the MD time by $\tau$
with the implicit leapfrog can then
be represented as:
\begin{align}
  T^{p,\tau/2}
  \cdot
  (T^{q,-\tau/2})^{-1}
  \cdot
  T^{q,\tau/2}
  \cdot
  (T^{p,-\tau/2})^{-1}.
\end{align}
Such a simplified viewpoint
enables us to easily write down
improved integrators.
For example,
the Omelyan integrator \cite{Omelyan_2002, TAKAISHI20006,PhysRevE.73.036706}
may be written in the non-separable case as:
\begin{align}
  &T^{p,\alpha\tau}
  \cdot
  (T^{q,-\alpha\tau})^{-1}
  \cdot
  T^{q,(1/2-\alpha)\tau}
  \cdot
  (T^{p,-(1/2-\alpha)\tau})^{-1}
   \nonumber\\
  &~~
\cdot
    T^{p,(1/2-\alpha)\tau}
  \cdot
  (T^{q,-(1/2-\alpha)\tau})^{-1}
  \cdot
  T^{q,\alpha\tau}
  \cdot
    (T^{p,-\alpha\tau})^{-1}.
    \label{eq:omelyan}
\end{align}

As is well known \cite{TANG199431,Kennedy:2012gk},
there exists
an exactly conserved shadow Hamiltonian $\tilde H$
for a given symplectic integrator.
For the implicit Omelyan~\eqref{eq:omelyan},
for example,
the shadow Hamiltonian has the formal expansion:
\begin{align}
  \tilde{H} &= H + \frac{\tau^2}{12} \Big[
   (6\alpha^2 -6\alpha + 1)
    \frac{\partial H}{\partial w_I}
    \frac{\partial^2 H}{\partial p_I \partial p_J}
    \frac{\partial H}{\partial w_J}
    \nonumber\\
  & + \frac{6\alpha - 1}{2}
      \frac{\partial H}{\partial p_I}
      \frac{\partial^2 H}{\partial w_I \partial w_J}
      \frac{\partial H}{\partial p_J}
      \nonumber\\
  & + (12\alpha^2 - 6\alpha + 1)
      \frac{\partial H}{\partial w_I}
      \frac{\partial^2 H}{\partial p_I \partial w_J}
      \frac{\partial H}{\partial p_J}
      \Big]  + \mathcal{O}(\tau^4).
\end{align}
Since we do not know the typical values of the 
derivatives
a prior, the parameter $\alpha$ is to be tuned manually
for a given system.

\section{Complications of non-separable 
Hamiltonians for group variables}
\label{sec:problem}

With the symplectic form for the $SU(N)$ variable \cite{Kennedy:1989ae}:
\begin{align}
  \omega_{SU(N)}(U,\pi) \equiv d( \pi_a \Theta_a ),
  \label{eq:symp_sun}
\end{align}
the continuous Hamiltonian equations can be written down as:
\begin{align}
  &\dot{U} U^{-1}
  = 
  iT_a \partial_{\pi_a} H,
 \label{eq:Hamiltonian_eq1}
  \\
  &\dot{\pi}_a 
  = 
  -D_a H - \pi_c f_{abc} \partial_{\pi_b}H,
  \label{eq:Hamiltonian_eq2}
\end{align}
under which 
$\dot{H}=0$ and $\dot{\omega}=0$.
Note here that $\Theta$ is not closed, and thus:
\begin{align}
  d\Theta_a = -\frac{1}{2}f_{bca} \,\Theta_b \wedge \Theta_c,
\end{align}
where $[T_a, T_b] \equiv i f_{abc} T_c$.

The problem for the case of a non-separable Hamiltonian~\eqref{eq:nonsepH}
arises for gauge theory
when we discretize the Hamiltonian equations~\eqref{eq:Hamiltonian_eq1} and \eqref{eq:Hamiltonian_eq2}.
As a simple example,
let us consider
applying the implicit leapfrog as follows:
\begin{widetext}
\begin{align}
  & \pi^{1/2}_a = \pi _a
    -
    \frac{\tau}{2}
    \big\{
    D_a H(U,\pi^{1/2})
    +
    \pi^{1/2}_c f_{abc} \partial_{\pi_b}H(U,\pi^{1/2})
    \big\}, \\
  & U^{1/2} = \exp\Big[
    \frac{i \tau}{2} T_a \partial_{\pi_a} H(U,\pi^{1/2})
    \Big] U, \label{eq:2nd}\\
  &U' = \exp\Big[
    \frac{i \tau}{2} T_a \partial_{\pi_a} H(U',\pi^{1/2})
    \Big] U^{1/2}, \\
  &\pi'_a = \pi^{1/2}_a
    -
    \frac{\tau}{2}
    \big\{
    D_a H(U',\pi^{1/2})
    +
    \pi^{1/2}_c f_{abc} \partial_{\pi_b}H(U',\pi^{1/2})
    \big\}.
\end{align}
\end{widetext}
To see the non-preservation of $\omega$,
we write the differential of the exponential map as:
\begin{align}
  iT_b A(X)_{ba}
  \equiv
  \partial_{X_a} e^{i X_c T_c}
  \cdot
  e^{ - iX_c T_c},
\end{align}
or equivalently:
\begin{align}
  A(X) &= \frac{\exp(i X_c F_c) 
  -
  \mathbb{1}_{N^2-1}}{i X_c F_c}
  \nonumber\\
  & =
  \sum_{k=0}^\infty \frac{1}{(k+1)!} 
  (i X_c F_c)^k,
\end{align}
where $(F_c)_{ab} \equiv if_{acb}$.
We have for the second step, Eq.~\eqref{eq:2nd}:
\begin{align}
  \Theta^{1/2}_b
  =
  \exp(i X_c F_c )_{ba} \Theta_a
  +
  A(X)_{ba} dX_a,
\end{align}
where $X_a \equiv \frac{\tau}{2} \partial_{\pi_a} H(U,\pi^{1/2})$.
The exponentiation
resulted in the nonlinear factors,
$\exp(i X_c F_c)$ and $A(X)$.

Note that for a trivial kernel, $G=\mathbb{1}_{N^2-1}$,
we have $X = (\tau/2) \pi^{1/2}$.
Because of the antisymmetry of $f_{abc}$,
the nonlinearity drops out in this case
when they are contracted with $\pi^{1/2}$:
$X_b \exp(i X_c F_c)_{ba} = X_a$
and
$X_b A(X)_{ba} = X_a$.
As a result,
the symplectic form is preserved at each step:
$\omega_{SU(N)}(U^{1/2},\pi^{1/2}) = \omega_{SU(N)}(U,\pi^{1/2}) = \omega_{SU(N)}(U,\pi)$.
On the other hand, for a non-separable Hamiltonian,
the preservation of $\omega_{SU(N)}$ is 
designed to hold for the paired update
of the first two steps
[see Eq.~\eqref{eq:conservation_unconstrained}].
Because of the complicated form of $X$,
the nonlinear terms do not simplify as in the former case,
and there are no counterterms from the first step
that cancel them.
Therefore, the naive application of the implicit leapfrog
to a non-separable Hamiltonian
in gauge theory generically violates  symplecticity.

To overcome the issue,
one may consider
introducing coordinate systems
instead of working in the 
invariant bases $\{\Theta_a\}$, $\{D_a\}$.
This turns out, however, not to be straightforward.
For simplicity,
let us take $SU(N=2) \simeq S^3$ as an example.
One may divide $S^3$ into two hyper-hemispherical patches
and introduce 
exponential coordinates
$q_{N/S}$ around the pole points $U_{N/S}$ (where N/S is an abbreviation for North/South):
\begin{align}
  U
  =
  \exp[i q_{N/S}^a T_a] U_{N/S}.
  \label{eq:coords}
\end{align}
By defining the momentum $p_a$ conjugate to $q^a$ ($N/S$ labels are suppressed):
\begin{align}
  p_a \equiv \pi_{b} A(q)_{ba},
  \label{eq:p_pi_map}
\end{align}
the symplectic form~\eqref{eq:symp_sun}
will have a flat form:
\begin{align}
  \omega_{SU(N)} = dp_a \wedge dq^a.
\end{align}
We can therefore apply
the implicit, symplectic integrators
for the unconstrained variables
in a given patch by taking
$(q,p)$ as the phase space variables.

The complication arises when
the trajectory steps over the boundary between the patches.
Note that 
preserving the symplecticity and 
the conservation of $H$
can be dealt with straightforwardly because
we know the transformation laws
for $(q^a, p_a)$ from
Eqs.~\eqref{eq:coords} and \eqref{eq:p_pi_map}.
However, it is difficult to determine
which coordinate system to use
in a reversible way.
In fact, for a finite $\tau$,
the force vector differs
depending on
the choice of the coordinates.
Accordingly,
tedious exception handling occurs
for the cases where
the update steps over
the boundary with the initially chosen coordinate system
but does not for the reversed sequence.
For a nontrivial kernel $G$,
the change of coordinates
on one link affects the determination for the other links,
giving rise to a global optimization problem.
We did not pursue this direction
since it was already difficult to maintain reversibility
for a small-size system in a systematic way.

\section{More on the exactness of the algorithm}
\label{sec:reversibility}

In this appendix, we elaborate on reversibility for the decomposition scheme developed in Sect.~\ref{sec:polar}.
Since the main concern will be in relation to the center ambiguity, we demonstrate the reversibility first for the update in the $U(1)$ direction and drop the $\Phi$-dependence:
\begin{align}
    \Omega'
    \equiv 
    e^{i\delta \omega} \Omega.
    \label{eq:omega'}
\end{align}
According to Eq.~\eqref{eq:dtheta}, we calculate:
\begin{align}
    \delta\theta
    =
    \frac{1}{N}
    \arg e^{iN\delta\omega}
    =
    \frac{1}{N}
    \floor{N \delta\omega},
    \label{eq:deltatheta_app}
\end{align}
where $\floor{\theta}$ is the projection of the angular variable $\theta$ onto the range $[-\pi,\pi)$.
We obtain:
\begin{align}
    U' 
    =
    e^{i(\delta \omega -\delta\theta)} U
    =
    e^{i
    (\delta \omega 
    -
    \frac{1}{N}
    \floor{N\delta\omega}
    )} U
    .
    \label{eq:u1_update}
\end{align}
The map is manifestly reversible, except for the case with $\floor{N \delta\omega} = -\pi$ to be dealt with later, because the reverse process corresponds to switching the sign of $\delta\omega$, for which Eq.~\eqref{eq:u1_update} still applies with the negative sign, correctly reproducing $(U,\theta)$ from $(U', \theta')$.
A general update of $\Omega$ can be given by a left multiplication of an $SU(N)$ matrix in Eq.~\eqref{eq:omega'}, which does not change the conclusion.

Given that our decomposition scheme provides a reversible update, the HMC remains exact.
In fact, as mentioned in Sect.~\ref{sec:rmhmc_gln}, a state of the Markov chain can be labeled as $(W,n)$ $(W \in GL(N, \mathbb{C}), n \in \mathbb{Z}_N)$.
Note however that the volume element in the $W$-space does not depend on $n$ for the same $W$.
The leapfrog integrators in the $W$-space guarantee the volume preservation for the extended phase space measure $(dW)(dp)$ thanks to the functional form of the force vector, irrespective of $n$ that the updated configuration will be labeled by.
The reversibility and the volume preservation then guarantee the exactness of the algorithm.

The exceptional case with $\floor{N \delta\omega} = -\pi$ is measure-zero, and thus it does not affect the algorithm under perfect real arithmetic.
Nevertheless, it may be of academic interest to show that the algorithm is exact even when the exceptional case is taken into account by using the skew detailed balance \cite{TURITSYN2011410}.
Let us write the transition probability from the state $(U,\theta,\Phi)$ to $(U',\theta',\Phi')$ as $P(U',\theta',\pi' \vert U,\theta,\Phi)$, in which the range of the angular variable is $[-\pi, \pi)$. 
We further introduce the transition probability $P_R(U',\theta',\Phi' \vert U,\theta,\Phi)$ that represents the same algorithm except that the range of the angular variable is replaced by $(-\pi, \pi]$.
Note that $P_R$ serves as the reverse algorithm of $P$ including the exceptional case by accommodating the edge value with the different sign. 
We then see that they satisfy the skew detailed balance relation:
\begin{align}
    &P(U',\theta',\Phi' \vert U,\theta,\Phi)
    \, e^{-S(U)-S_0(\Phi)} \nonumber\\
    &=
    P_R( U,\theta,\Phi \vert U',\theta',\Phi')
    \, e^{-S(U')-S_0(\Phi')},
\end{align}
which shows that the Boltzmann factor is the fixed point of the transition matrix $P$ as we see by integrating over the triplet $(U,\theta,\Phi)$.
The algorithm is therefore exact.

We finally comment on the discontinuous jump of $(U',\theta')$ at $\delta\omega = \pi/N$ which we encounter when we continuously increase $\delta \omega$ in Eq.~\eqref{eq:deltatheta_app}.
Such a jump generically implies a significant deviation from the continuous time trajectory, leading to a non-energy conservation.
While the actual probability of encountering the discontinuity can be system-dependent, we can argue that it only happens when the acceptance rate would be low in the first place due to the appearance of the large force $F$.
In fact, for the worst case where $F$ is aligned with the $U(1)$ direction, a nontrivial jump occurs when we have the force with a sizable magnitude for the single link variable: $\tau F \sim O(1/N)$, where $\tau F$ should usually be suppressed by a rational order of the volume.
At least for small $N$, such a sizable update would mean a mistuning of $\tau$ that leads to a small acceptance rate.
Conversely, in practical calculations with large volumes, 
the discontinuous jump rarely occurs and hardly affects the acceptance rate. 
We emphasize that the algorithm is exact even with the discontinuous jumps as discussed in the first half of the appendix.

\section{Alternative decomposition \\
without the \texorpdfstring{$\mathbb{Z}_{N}$}{ZN} ambiguity}
\label{sec:another_decomp}

In this appendix, 
we consider an alternative decomposition:
\begin{align}
    W = \Phi 
    \exp({i \theta T_0}) U,
    \quad
    T_0 \equiv {\rm diag}(1,0,\cdots,0),
    \label{eq:alternative_decomposition}
\end{align}
and show that 
the $\mathbb{Z}_N$ 
ambiguity 
mentioned in Sec.~\ref{sec:polar}
can be removed. 
The argument holds for any hermitian matrix 
$T_0$ with the same
eigenvalues and their geometric multiplicity as those of the $T_0$ 
defined by Eq.~\eqref{eq:alternative_decomposition}.

As in the main text, we first perform the polar decomposition~\eqref{eq:polar_VQ}
to obtain $\Omega$ and $\Phi$.
We then set $\theta$ and $U$ as:
\begin{align}
    \theta \equiv \arg \det \Omega,
    \quad
    U \equiv \exp({-i \theta T_0}) \Omega.
    \label{eq:alternative_theta_U}
\end{align}
The Jacobian matrix $J(U,\theta,\Phi)$
will be modified as:
\begin{widetext}
\begin{align}
J(U, \theta, \Phi)
=
\left[
    \begin{array}{c c}
      - {\rm Im}\,[\Phi \exp({i \theta T_0}) T_a U]_{jk}
      &
        {\rm Re}\,[\Phi \exp({i \theta T_0}) T_a U]_{jk}
      \\
      - {\rm Im}\,[e^{i \theta} (\Phi T_0 U)_{jk}]
      &
        {\rm Re}\,[e^{i \theta} (\Phi T_0 U)_{jk}]
      \\
      {\rm Re}\,[ T_a \exp({i \theta T_0}) U]_{jk}
      &
        {\rm Im}\,[ T_a \exp({i \theta T_0}) U]_{jk}
      \\
      {\rm Re}\,[\exp({i \theta T_0})U]_{jk}
      &
        {\rm Im}\,[\exp({i \theta T_0}) U]_{jk}
    \end{array}
    \right].
\end{align}
\end{widetext}

Compared to Eq.~\eqref{eq:solutions_ZN},
with this decomposition,
$(U,\theta)$ can be simply given from $\Omega$
by Eq.~\eqref{eq:alternative_theta_U}
and there are no multiple solutions with $\theta$ in the range
$-\pi\leq \theta <\pi$.
The removal of the 
$\mathbb{Z}_N$ 
ambiguity can be understood as a breaking of the connection between
the elements of the center 
$\mathbb{Z}_N$ 
and the angle $\theta$.  Specifically, a right multiplication of
$\exp(i\theta T_0)$ 
by an element of $\mathbb{Z}_N$,
$e^{2\pi i n/N}\mathbb{1}_N$,
cannot be absorbed 
by a shift of $\theta$.

With the decomposition given in the main text, the $\mathbb{Z}_N$ ambiguity demands special care when $W$ is treated independently from the variables $(U,\theta,\Phi)$. This includes when we perform a hot start and when we save a configuration in terms of the $W$ variable 
(see Sec.~\ref{sec:rmhmc_gln}).
The removal of the $\mathbb{Z}_N$ ambiguity simplifies the algorithm
in this regard.

We provide an implementation 
of the algorithm 
with the decomposition~\eqref{eq:alternative_decomposition} 
in the Grid
Python Toolkit (GPT) \cite{GPT2}, which uses Grid \cite{Boyle:2016lbp,Yamaguchi:2022feu} for performance portability.

\section{Reduced Hamiltonian equations}
\label{sec:reduced}

In this appendix, we derive
the reduced Hamiltonian equations 
in the variables $(U, \theta, \Phi)$
in the continuous time limit.

It is convenient to write the decomposed bases 
collectively as:
\begin{align}
  (\Theta^A) &\equiv (\Theta_a, d\theta, d\phi_a, d\phi_0),
  \\
  (D_A) &\equiv (D_a, \partial_\theta, \partial_{\phi_a}, \partial_{\phi_0}).
\end{align}
We distinguish the upper and lower indices in the extended bases.
The relation between the Jacobian matrix
and the metric tensor~\eqref{eq:metric_tensor} is:
\begin{align}
  dw^I
  = \Theta^A \tensor{J}{_A^I}(w),
  \quad
  \textbf{g}
   =
   g_{AB}(w)
   \Theta^A 
   \Theta^B
\end{align}
with 
\begin{align}
g_{AB}(w) 
  &\equiv \delta_{IJ}
  \tensor{J}{_A^I}(w) \tensor{J}{_B^J}(w).
\end{align}
Defining the pairing between a 
tangent vector (or simply a differential operator) $D$ 
and a one-form $df$:
\begin{align}
    \langle D, df \rangle
    \equiv D f,
\end{align}
we find $\langle D_A, \Theta^B\rangle = \delta_A^B$.
In particular, 
\begin{align}
    &\langle D_a , \Theta_b \rangle
    =
    -i\tr{[T_b D_aU U^{-1}]}
    =
    \delta_{ab},
    \\
    &\langle \partial_{\phi_a}, d\phi_b \rangle 
    = 
\partial_{\phi_a} \phi_b = \delta_{ab},
\end{align}
where we used 
$\Theta_a = -i\tr{[T_a dU U^{-1}]}$.

We first note that,
for an $SU(N)$ variable,
the Hamiltonian equations
with the symplectic form~\eqref{eq:symp_sun} 
can be written as:\footnote{%
In general,
given a basis
of the tangent space of the phase space,
$\{D_\mu\}$,
and the dual basis $\{\Theta^\mu\}$,
we have the Hamiltonian 
vector field 
(see, e.g., Ref.~\cite{Kennedy:2012gk}):
\begin{align}
    \frac{d}{dt} \equiv
    \omega^{\mu \nu} 
    (D_\nu H) D_\mu,
\end{align}
where 
\begin{align}
    \omega \equiv \frac{1}{2} 
    \omega_{\mu\nu} 
    \Theta^\mu \wedge \Theta^\nu,
    \quad
    (\omega^{\mu \nu}) 
    \equiv 
    (\omega_{\mu \nu})^{-1},
\end{align}
and $\omega_{\mu\nu} = -\omega_{\nu\mu}$.
For $\omega=\omega_{SU(N)}$, 
with
$\{D_\mu\} = \{D_a, \partial_{\pi_a}\}$
and
$\{\Theta^\mu\} = 
\{\Theta_a, d{\pi_a}\}$,
Eqs.~\eqref{eq:u_HE_app} and~\eqref{eq:u_HE_app2}  can be derived
by noting that:
\begin{align}
    (\omega_{\mu\nu})
    &=
    \left(
    \begin{array}{cc}
         -\pi_c f_{abc} & -\delta_{ab} \\
         \delta_{ab}   & O
    \end{array}
    \right),
    \\
    (\omega^{\mu\nu})
    &=
    \left(
    \begin{array}{cc}
         O & \delta_{ab} \\
         -\delta_{ab} & -\pi_c f_{abc}
    \end{array}
    \right).
\end{align}
}
\begin{align}
  &\Big\langle \frac{d}{dt}, \Theta_a \Big\rangle
  =
  \partial_{\pi_a} H,
  \label{eq:u_HE_app}
  \\
  &\frac{d \pi_a}{dt}
  =
  - D_a H - \pi_c \tensor{f}{_{abc}} \partial_{\pi_b} H,
  \label{eq:u_HE_app2}
\end{align}
where $t$ is the MD time.
Equation~\eqref{eq:u_HE_app} is equivalent to Eq.~\eqref{eq:Hamiltonian_eq1}
because:
\begin{align}
    iT_a \Big\langle \frac{d}{dt}, \Theta_a \Big\rangle
    =
    \Big\langle \frac{d}{dt}, \Theta \Big\rangle
    =
    \dot{U} U^{-1}
    =
    iT_a \partial_{\pi_a} H,
\end{align}
where $\dot{U}\equiv dU/dt$.

Next, in the extended system, we have the symplectic form:
\begin{align}
  \omega 
  =
  dp_I \wedge dw^I
  =
  d(\pi_A \Theta^A),
\end{align}
where 
\begin{align}
  \pi_A \equiv \tensor{J}{_A^I}(w) \, p_I.
\end{align}
As in Sec.~\ref{sec:discussion},
we write $(\pi_A) = (\pi_a, \pi_\theta, \rho_a, \rho_0)^T$.
The Hamiltonian equations can be written in two different ways:
\begin{align}
  \frac{d w^I}{dt} = \partial_{p_I}H,
  \quad
  \frac{d p_I}{dt} = -\partial_{w^I}H,
  \label{eq:w_basis_HE_app}
\end{align}
which are in the canonical form, and
\begin{align}
  &\Big\langle \frac{d}{dt}, \Theta^A \Big\rangle
  =
  \partial_{\pi_A} H,
  \label{eq:u_basis_HE_app0}
  \\
  &\frac{d \pi_A}{dt}
  =
  - D_A H - \pi_C \tensor{f}{_{AB}^C} \partial_{\pi_B} H,
  \label{eq:u_basis_HE_app}
\end{align}
analogously to Eqs.~\eqref{eq:u_HE_app}
and \eqref{eq:u_HE_app2},
where 
\begin{align}
    d\Theta^A \equiv - (1/2) \Theta^B \wedge \Theta^C \tensor{f}{_{BC}^A}.
\end{align}

The continuous time limit
of the MD update
described by
Eqs.~\eqref{eq:p_half}--\eqref{eq:p_prime}
is Eq.~\eqref{eq:w_basis_HE_app}.
Its equivalent description in 
the $(U,\theta,\Phi)$-basis
is, therefore, Eqs.~\eqref{eq:u_basis_HE_app0} and \eqref{eq:u_basis_HE_app}.
Note that
$\Theta^A$ are closed forms except for $\Theta_a$.
Accordingly, 
the structure constants $\tensor{f}{_{AB}^C}$ are only nontrivial
within the $SU(N)$ sector,
and they reduce to $f_{abc}$ in this sector.
For a block-diagonal
kernel $K(w)$
as given in Eq.~\eqref{eq:block_diag_kernel},
the time evolution of the
$(U,\theta,\Phi)$ variables
factorize in Eqs.~\eqref{eq:u_basis_HE_app0}
and~\eqref{eq:u_basis_HE_app}.
We thus see
that the reduced Hamiltonian
equations for the 
physical variable $U$
are
exactly Eqs.~\eqref{eq:u_HE_app}
and~\eqref{eq:u_HE_app2}.
One can further derive the reduced equations for $\theta$,
$\phi_a$ and $\phi_0$,
which have the canonical form. For example:
\begin{align}
  \frac{d \phi_a}{dt} = \partial_{\rho_a} H,
  \quad
  \frac{d \rho_a}{dt} = - \partial_{\phi_a} H.
\end{align}

\newpage

\bibliography{ref}

\end{document}